\documentclass{cargese}\usepackage{graphicx}
\let\footnote\savefootnote
\let\footnotetext\savefootnotetext 
\let\footnoterule\savefootnoterule 
\normallatexbib
\newcommand{\lag}{{\cal L}}

\newcommand{\Str}{{\mathbf {STr}}}
\newcommand{\Tr}{{\mathbf {Tr}}}

\newcommand{\T}{{\mathbf T}}
\newcommand{\F}{{\mathbf F}}

\newcommand{\D}{{\mathbf D}}

\begin{document}
\articletitle[On the non-abelian Born-Infeld action]{On the
non-abelian Born-Infeld action{\footnote{LPTENS-99/34}} } 
\author{Pascal Bain}
\affil{Laboratoire de Physique Th\'eorique de l'\'Ecole Normale
Sup\'erieure{\footnote{UMR 8549 du Centre National de la Recherche
Scientifique et de l'\'Ecole Normale Sup\'erieure.}} \\
24 rue Lhomond, F-75231 Paris Cedex 65.
}         

\begin{abstract}
We discuss some aspects of the generalization of the
Born-Infeld action to non-abelian gauge groups and show
how the discrepancy between Tseytlin's symmetrized trace proposal and
string theory can be corrected at order $F^6$. We also comment on the
possible quadratic order fermionic terms.
\end{abstract}

While the talk I gave at the Gong Show session of Carg\`ese' 99 Summer
School was
devoted to an outline of the main results of the paper ``Curvature
terms in D-brane actions and their M-theory origin'' \cite{bbg}, this  
note will be focused on another very interesting but
not yet settled aspect of the D-brane effective action~: the
generalization of the Dirac-Born-Infeld (DBI) action to non-abelian
gauge groups.

The massless degrees of freedom of an isolated D-brane are its
transverse coordinates, a world-volume $U(1)$ gauge field, and the
associated fermions \cite{review}.
Its effective action contains a  
Dirac-Born-Infeld (DBI) (see \cite{t1} for a recent review of the
DBI action) and a Wess-Zumino (WZ) piece.  While the WZ part,
describing  the interactions of the D-brane with 
Ramond-Ramond field backgrounds, can be determined by anomaly
cancellation and is therefore believed to be exact,
the DBI piece is only reliable 
for small space-time curvatures, and in the low-acceleration
regime. Indeed, for non constant gauge field strength, the DBI action
is corrected by derivative terms \cite{t1, at}. 

When $N$ identical D-branes coincide, the transverse coordinates and
world-volume gauge fields become $N \times N$ unitary matrices. The
non-abelian generalization of the DBI action describing
this dynamics has been the subject of several recent papers. 
In the low-energy non-relativistic limit one must recover
the usual quadratic action of dimensionally-reduced super Yang-Mills.
The question is whether there exist less-trivial truncations, taking
into account relativistic effects in a controlled manner. 
However, since $[\D_m, \D_n]\F_{pq} = [\F_{mn}, \F_{pq}]$ for a
non-abelian group, the part of the action which
depends on the field strength but not on its covariant derivative
is ambiguous, contrary to the abelian case. 
One simple proposal for the  flat space-time action has been put
forward by Tseytlin \cite{t2}~: 
\begin{equation}\label{symtr}
\lag = \Str \sqrt{-\det(\eta + 2 \pi\alpha' \F)} 
\end{equation}
The prescription is to expand the 
Born-Infeld action as a formal power series, symmetrize
under all permutations of the (non-abelian) field strengths and then
take the trace.  The resulting action captures several features of
the full string theory.

First, since $\Str(t_8 \F^4) = \Tr(t_8 \F^4)$, it reproduces the already
well-known $F^4$ which can be 
obtained from a direct four-point function on the disk diagram
\cite{gwt} or by a two-loop $\beta$--function for the non-abelian
background gauge field in open string theory \cite{bp}.

Second, in four dimensions, it reduces to the quadratic Yang-Mills term 
for an (anti-)self-dual field strength as pointed out in \cite{b} where
BPS states of the non-abelian BI action were investigated.

Since it seems out of reach to test this proposal to higher
orders in $\alpha^\prime$ by a direct string theory computation -- it
would involve a six-point open string amplitude on 
the disk or a four-loop $\beta$ function! -- one must use indirect
arguments.  Hashimoto and Taylor have found a configuration \cite{ht}
which exhibits disagreements with the symmetrized trace prescription~:  
they considered a constant magnetic background field \cite{acny} or,
in the dual string picture, a pair of intersecting rotated 
D2-branes living on a torus. To the background considered in
\cite{ht}, it is possible to add a $U(1)$ part which corresponds to
the simultaneous rotation of the two branes in the dual picture. We
expect that the spectrum still only depends on the relative
orientation of the branes. This abelian gauge field imposes further
constraints on the form of the action.

For an (anti-)self-dual background, the spectrum obtained with
the symmetrized trace prescription exactly matches the string theory
spectrum, a result not fully realized in \cite{ht}. 
For more general constant fields, both results agree up to order
${(\alpha^\prime)}^4$, which is consistent with the fact that the
symmetrized trace is the correct answer for $F^4$ terms, as 
already pointed out. The discrepancy appears at the sixth order and
can be solved at this order by adding terms with 
commutators of the fields strength $\F = F^a \T^a$ to the action
(\ref{symtr}).   
For simplicity, we have only considered terms with two
commutators, but even with this restriction, the test does not
completly fix the action. If we define the tensors
\begin{eqnarray*}
{C^\pm}^{a_1 a_2 a_3 a_4 a_5 a_6} &= 
& \Tr(\T^{a_1}\T^{a_2}\{[\T^{a_3},\T^{a_4}],[\T^{a_5},\T^{a_6}]\})/4  \\
&\pm&  \Tr(\T^{\{a_1,}[\T^{a_3},\T^{a_4}]\T^{a_2\}}[\T^{a_5},\T^{a_6}])/4  \\
\end{eqnarray*} 
and the contractions 
\begin{eqnarray*}
V_1^\pm &=& (2\pi\alpha^\prime)^6 F^{a_1}_{mn} F^{a_2}_{np} F^{a_3}_{pq} F^{a_4}_{rm} F^{a_5}_{qs}
F^{a_6}_{sr} {C^\pm}^{a_1 a_2 a_3 a_4 a_5 a_6} \\
V_2^\pm &=& (2\pi\alpha^\prime)^6 F^{a_1}_{mn} F^{a_2}_{pq}
F^{a_3}_{nr} F^{a_4}_{sm} F^{a_5}_{rp} 
F^{a_6}_{qs} {C^\pm}^{a_1 a_2 a_3 a_4 a_5 a_6} \\
V_3^\pm &=& (2\pi\alpha^\prime)^6 F^{a_1}_{mn} F^{a_2}_{pq}
F^{a_3}_{np} F^{a_4}_{rs} F^{a_5}_{qm} 
F^{a_6}_{sr} {C^\pm}^{a_1 a_2 a_3 a_4 a_5 a_6} \\
V_4^\pm &=& (2\pi\alpha^\prime)^6 F^{a_1}_{mn} F^{a_2}_{pq}
F^{a_3}_{nm} F^{a_4}_{qr} F^{a_5}_{sp} 
F^{a_6}_{rs} {C^\pm}^{a_1 a_2 a_3 a_4 a_5 a_6} \\
V_5^\pm &=& (2\pi\alpha^\prime)^6 F^{a_1}_{mn} F^{a_2}_{pq}
F^{a_3}_{nm} F^{a_4}_{rs} F^{a_5}_{qp} 
F^{a_6}_{sr} {C^\pm}^{a_1 a_2 a_3 a_4 a_5 a_6}
\end{eqnarray*}
the string theory spectrum is reproduced at order six in
$\alpha^\prime$ when we add
$$\frac{1}{24}(V_1^+ - V_2^+ - \frac{1}{2} V_3^+ - V_4^+ + \frac{1}{4}
V_5^+) +
 \frac{1}{360}(V_1^- - V_2^- - \frac{1}{2} V_3^- - V_4^- + \frac{1}{4}
V_5^-)$$
to the action (\ref{symtr}). We have also checked that for an
(anti-)self-dual field strength, these additional terms vanish and,
then, the effective action reduces to the linear Yang-Mills action.

This situation seems somewhat similar to the problem studied by
Douglas, Kato and Ooguri \cite{dko} where they demonstrate the
necessity to add commutators terms to the symmetrized trace in their   
D-K\"ahler potential (at sixth order) to reproduce the mass-shell
conditions. 

\begin{figure}[t]
\begin{center}
\includegraphics[width=8cm]{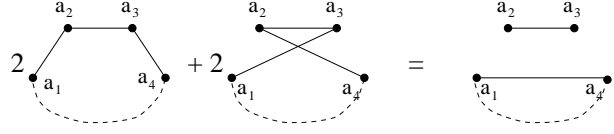}
\end{center}
\caption{Relation between diagrams for $F=\ ^\star F$.}\label{fig:sd}
\end{figure}

A more exhaustive and systematic investigation, combining  
the self-duality and Hashimoto-Taylor's constraints, would be
desirable. For this purpose, it seems convenient to use the following
diagramatic representation to enumerate all the possible terms
appearing at a given order $(\alpha^\prime)^n$~: the idea is to put
each $\F$ field at a vertex of a polygon of degree $n$ and to symbolize
each Lorentz contraction by a line joining these points. 
\begin{figure}[b]
\begin{center}
\includegraphics[width=8cm]{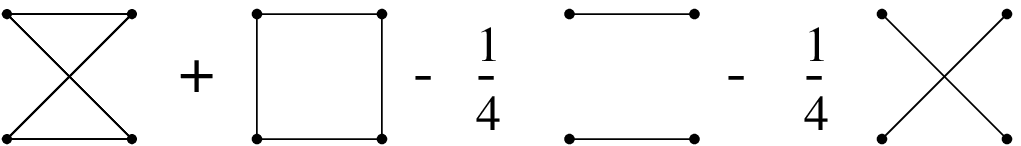}
\end{center}
\caption{Diagramatic representation of $3 \  \Str (\F^4) $.}\
\label{fig:f4s} 
\end{figure}
The trace over
the gauge group is taken in the standard order on the polygon.
For example, the $\Str(\F^4)$ terms are given in figure (\ref{fig:f4s}).
Notice that in the symmetrized trace, the weight of each term for a given
set of Lorentz contractions are just the
number of non-superimposable diagrams under rotations preserving the
polygon vertices~: in this way, in figure (\ref{fig:f4s}), the first
and third 
diagrams have weight two and the second and fourth weight one. Using
this representation, one can easily 
enumerate the possible terms at order $(\alpha^\prime)^6$; there are
exactly twenty eight inequivalent diagrams~: five of the ``$(F^2)^3$''
type, nine ``$(F^2)(F^4)$'' and fourteen ``$F^6$''.
Moreover, for a self-dual field strength, we have the
relation~:     
$$2(\F_{mp}\F_{pn} + \F_{np}\F_{pm}) = \eta_{mn} \F_{pq}\F_{qp}$$
which can be translated into the diagrammatic equation of figure
(\ref{fig:sd}). 
Since for a self-dual field strength higher-order terms should
vanish \cite{b}, this implies further constraints on the possible form
of the action. 
However, using this rule, it is quite simple to analyse these
constraints; for example, it is easy to verify that the $F^4$
term in figure (\ref{fig:f4s}) vanishes for a self-dual $F$.

\ 

Let us conclude with some remarks~: first, the effective action should
also contain fermionic terms.
Aganagic, Popescu and Schwarz have proposed a supersymmetric
generalization of the abelian BI action \cite{aps}. 
To study its generalization to non-abelian groups, we can use
a test analoguous to the one developed for the gauge field in \cite{ht}. 
From the dual string theory, one expect, for fermions in a
constant gauge field, a Landau spectrum of the form~: 
\begin{eqnarray*}
2\Sigma_{12}.\arctan(2\pi\alpha'F_{12}) +
2\Sigma_{34}.\arctan(2\pi\alpha'F_{34})
\end{eqnarray*} 
with $\Sigma_{ij}=[\Gamma_i,\Gamma_j]$.
Such spectrum can be obtained from the Lagrangian~:
\begin{eqnarray*}
\lag_2= \Str \left[\overline{\lambda'}{(\eta +
2 \pi\alpha' \F)^{-1}_S}^{\mu \nu} \Gamma_\mu \D_\nu\lambda' \right].
\end{eqnarray*}

Finally, Seiberg and Witten \cite{sw} have recently proved that the standard
abelian BI action is equivalent, up to terms involving derivatives of
$F$, to the abelian BI for a non-commutative gauge field by a field
redefinition. This result should still hold for the non-abelian
theory. It would be interesting to see what the symmetrized trace
prescription on one side gives on the other side. Moreover, in the
limit $\alpha^\prime \rightarrow 0$ considered in \cite{sw}, the
action on the 
non-commutative side reduces to the non-commutative Yang-Mills
Lagrangian whereas it remains non-trivial on the commutative
side. Since the generalization of this abelian NCYM to the
non-abelian case is trivial, one can ask how this discussion can be
extend and give some insights on the commutative non-abelian side.

\begin{acknowledgments}
I would like to thank the organizers of the Carg\`ese'99 ASI for a very nice
school and for financial support. It is a pleasure to thank C. Bachas
for collaboration on this subject. 
\end{acknowledgments}

\end{document}